\documentclass{article}
\usepackage{eurosym}
\usepackage{amssymb}
\usepackage{amsmath}
\usepackage{cite}
\usepackage{graphicx}
\usepackage{epsf}

\begin{document}

\title{Lie symmetries and similarity solutions for rotating shallow water}
\author{Andronikos Paliathanasis\thanks{%
Email: anpaliat@phys.uoa.gr} \\
{\ \textit{Instituto de Ciencias F\'{\i}sicas y Matem\'{a}ticas, Universidad
Austral de Chile, Valdivia, Chile}} \\
{\ \textit{Institute of Systems Science, Durban University of Technology }}\\
{\ \textit{PO Box 1334, Durban 4000, Republic of South Africa}}}
\maketitle

\begin{abstract}
We study a nonlinear system of partial differential equations which describe
rotating shallow water with an arbitrary constant polytropic index $\gamma $
for the fluid. In our analysis we apply the theory of symmetries for
differential equations and we determine that the system of our study is
invariant under a five dimensional Lie algebra. The admitted Lie symmetries
form the $\left\{ 2A_{1}\oplus _{s}2A_{1}\right\} \oplus _{s}A_{1}$ Lie
algebra for $\gamma \neq 1$ and $2A_{1}\oplus _{s}3A_{1}$ for $\gamma =1$.
The application of the Lie symmetries is performed with the derivation of
the corresponding zero-order Lie invariants which applied to reduce the
system of partial differential equations into integrable systems of ordinary
differential equations. For all the possible reductions the algebraic or
closed-form solutions are presented. Travel-wave and scaling solutions are
also determined.

\end{abstract}

\section{Introduction}

\label{sec1}

Lie symmetries is an essential tool for the study of nonlinear differential
equations. The main characteristic of the Lie symmetry analysis is that
invariant surfaces, in the space where the parameters of the nonlinear
differential equation evolve, are determined which can be used to performed
an extended analysis of the nonlinear differential equation \cite%
{Ovsi,harrison,refA1,ser1,ser2,nuc1,nuc2}, construct conservation laws \cite%
{noeth,nibr,sarlet} and when it is feasible to determine solutions of the
differential equation \cite{olver,kumei,ibra,serb}. In applied mathematics
Lie symmetries cover a wide range of applications from physics, biology,
financial mathematics and many others for instance see \cite%
{app1,app2,app3,app4,ap11,ap12,app5,app6,app7,app8,app9,app10,ap13,ap14} and
references therein.

In this work, we interest on the application of Lie's theory on an system of
partial differential equations (PDEs) describe one-dimensional rotating
shallow water phenomena. The system of our consideration expressed in
Lagrangian coordinates is \cite{ref1}
\begin{eqnarray}
h_{t}+h^{2}u_{x} &=&0  \label{sw.01} \\
u_{t}+h^{\gamma -1}h_{x}-v &=&0  \label{sw.02} \\
v_{t}+u &=&0  \label{sw.03}
\end{eqnarray}%
where $h=h\left( t,x\right) $ denotes the height of the fluid surface, $%
u=u\left( t,x\right) $ denotes the velocity component in the $x$-direction
and $v=v\left( t,x\right) $ is the other horizontal velocity component that
is in the direction orhtogonal to the $x-$direction \cite{ref1}. Parameter $%
\gamma $ is the polytropic parameter of the fluid, where in this work is
assumed to be $0\leq \gamma .$ The system (\ref{sw.01})-(\ref{sw.03}) is
important for the study of atmospheric phenomena like geostrophic adjustment
and zonal jets, for more details of the physical properties of the above
system we refer the reader in \cite{ref2,ref3,ref4} and references therein.

The application of Lie symmetries in shallow-water theory is not new.
Indeed, there are various studies in the literature \cite%
{swr02,swr03,swr04,swr05,swr06,swr07} which has been provide important
results with special physical interest. Recently, a detailed study of the
nonlocal symmetries for a variable coefficient shallow water equation
performed in \cite{swr08}. However, the majority of these studies are for
the case where the fluid has a specific polytropic exponent $\gamma $, or
the shallow water equations describe non-rotating phenomena. The plan of the
paper is as follows.

In Section \ref{sec2} we present the basic properties and definitions of Lie
symmetry analysis which is the main mathematical tool for our analysis. The
main results of this work are presented in Section \ref{sec3}. More
specifically, we reduce the system (\ref{sw.01})-(\ref{sw.03}) into two
equations for the variables $h$ and $v$. We derive that the latter system of
two PDEs admits five Lie point symmetries and we study all the possible
reductions in ordinary differential equations \ (ODEs) with the use of
zero-order Lie invariants. We find that the reduce systems can be solve
explicitly and we derive the algebraic solution or closed-form solutions for
every possible reduction and every value of the parameter $\gamma $. The
latter result is important because it shows how powerful is the method of
Lie symmetry analysis for the study of shallow-water phenomena to prove the
existence of solutions for the model of our study. Emphasis is given on the
travel-wave and scaling solutions. Finally our discussion and conclusions
are presented in\ Section \ref{sec4}.

\section{Preliminaries}

\label{sec2}

In this Section we briefly discuss the basic definitions and main steps for
the determination of Lie point symmetries for differential equations.

Consider a system of PDEs
\begin{equation}
H^{A}\left( y^{i},u^{A},u_{i}^{A},...\right) =0,  \label{de.01}
\end{equation}
where $u^{A}~$denotes the dependent variables, $y^{i}$ are the indepedent
variables$~$and $u_{i}^{A}=\frac{\partial u^{A}}{\partial y^{i}}$.

We assume the one-parameter point transformation (1PPT) in the space of the
independent and dependent variables
\begin{eqnarray}
\bar{y}^{i} &=&y^{i}\left( y^{j},u^{B};\varepsilon \right) ,  \label{de.02}
\\
\bar{u}^{A} &=&u^{A}\left( y^{j},u^{B};\varepsilon \right) ,  \label{de.03}
\end{eqnarray}%
in which $\varepsilon $ is an infinitesimal parameter, the differential
equation (\ref{de.01}) remain invariant if and only if
\begin{equation}
\bar{H}^{A}\left( \bar{y}^{i},\bar{u}^{A},...;\varepsilon \right)
=H^{A}\left( y^{i},u^{A},...\right) ,  \label{ls.04}
\end{equation}%
or equivalently \cite{kumei}%
\begin{equation}
\lim_{\varepsilon \rightarrow 0}\frac{\bar{H}^{A}\left( \bar{y}^{i},\bar{u}%
^{A},...;\varepsilon \right) -H^{A}\left( y^{i},u^{A},...\right) }{%
\varepsilon }=0.  \label{ls.05}
\end{equation}

The latter conditions is expressed%
\begin{equation}
\mathcal{L}_{X}\left( H^{A}\right) =0,  \label{ls.05a}
\end{equation}%
where~$\mathcal{L}$ denotes the Lie derivative with respect the vector field
$X^{\left[ n\right] }$ which is the $n$th-extension of generator $X~$of the
infinitesimal transformation (\ref{de.02}), (\ref{de.03}) in the jet space $%
\left\{ y^{i},u^{A},u_{,i}^{A},...\right\} $
\begin{equation}
X^{\left[ n\right] }=X+\eta ^{\left[ 1\right] }\partial
_{u_{i}^{A}}+...+\eta ^{\left[ n\right] }\partial
_{u_{i_{i}i_{j}...i_{n}}^{A}},  \label{ls.06}
\end{equation}%
with generator
\begin{equation}
X=\frac{\partial y}{\partial \varepsilon }\partial _{x}+\frac{\partial u^{A}%
}{\partial \varepsilon }\partial _{u^{A}},  \label{de.07}
\end{equation}%
and%
\begin{equation}
\eta ^{\left[ n\right] }=D_{i}\eta ^{\left[ n-1\right]
}-u_{i_{1}i_{2}...i_{n-1}}D_{i}\left( \frac{\partial y}{\partial \varepsilon
}\right) ~,~i\succeq 1.  \label{de.08}
\end{equation}

When condition (\ref{ls.05a}) is satisfied for a specific 1PPT, the vector
field $X$ is called a Lie point symmetry for the system of PDEs (\ref{de.01}%
). For an unknown 1PPT, in order to specify the generators $X$ which are Lie
point symmetries for a given differential equation, from the symmetry
condition (\ref{ls.05a}) we specify a system of PDEs with dependent
variables the components of the generator $X$. The solution of the latter
system provides the generic symmetry vector and the number of independent
solutions give the number of indepedent vector field and the dimension of
the admitted Lie algebra.

\section{Lie symmetry analysis}

\label{sec3}

We write the system (\ref{sw.01})-(\ref{sw.03}) as two second-order PDEs%
\begin{eqnarray}
v_{tx}-h^{-2}h_{t} &=&0,  \label{sw.04} \\
v_{tt}-h^{\gamma -1}h_{x}+v &=&0  \label{sw.05}
\end{eqnarray}%
while the application of Lie's theory provides a five dimensional Lie
algebra consists by the following vector fields%
\begin{eqnarray*}
X_{1} &=&\partial _{t}~,~X_{2}=\partial _{x}~,~ \\
X_{3} &=&\cos \left( t\right) \partial _{v}~,~X_{4}=\sin \left( t\right)
\partial _{v} \\
X_{5} &=&\left( \gamma +1\right) x\partial _{x}+\left( \gamma -1\right)
v\partial _{v}+2h\partial _{h}
\end{eqnarray*}%
In table \ref{tabl1} the commutators of the Lie symmetries are presented.
Consequently, from table \ref{tabl1} we can refer that the admitted Lie
algebra is the $\left\{ 2A_{1}\oplus _{s}2A_{1}\right\} \oplus _{s}A_{1}~$in
the Morozov-Mubarakzyanov Classification Scheme \cite%
{Morozov58a,Mubarakzyanov63a,Mubarakzyanov63b,Mubarakzyanov63c}. However, in
the limit where $\gamma =1$, the admitted Lie algebra is the $2A_{1}\oplus
_{s}3A_{1}$.

In order to continue with the application of the Lie point symmetries it is
important to determine the one-dimensional optimal system and invariants
\cite{olverb}. In order to do that the adjoint representations should be
calculated. By definition, for every basis of the Lie symmetries $X_{i}$,
the adjoint representation is given by the following expression
\begin{equation}
Ad\left( \exp \left( \varepsilon X_{i}\right) \right)
X_{j}=X_{j}-\varepsilon \left[ X_{i},X_{j}\right] +\frac{1}{2}\varepsilon
^{2}\left[ X_{i},\left[ X_{i},X_{j}\right] \right] +...~\text{.}
\label{ad.01}
\end{equation}%
For the admitted Lie point symmetries of the system (\ref{sw.04}), (\ref%
{sw.05}) the adjoint representation are given in \ref{tabl2}. In order to
find the optimal system we consider the generic symmetry vector
\begin{equation}
\mathbf{X}=a_{1}X_{1}+a_{2}X_{2}+a_{3}X_{3}+a_{4}X_{4}+a_{5}X_{5}
\label{ad.02}
\end{equation}
and we find the equivalent vectors by considering the adjoint
representation. At this point it is important to mention that the adjoint
action admits two invariant functions, the $\phi _{1}\left( a_{i}\right)
=a_{1}$ and $\phi _{2}\left( a_{i}\right) =a_{5}~$\cite{opt1}. The
invariants can be used to simplify the calculations on the derivation of the
optimal system. Indeed we have to consider the cases $a_{1}a_{2}\neq 0$ and $%
a_{1}a_{2}=0$.

Case 1: For  $a_{1}a_{5}\neq 0$, we have that%
\begin{equation*}
\mathbf{X}^{\prime }=Ad\left( \exp \left( \varepsilon _{4}X_{4}\right)
\right) Ad\left( \exp \left( \varepsilon _{3}X_{3}\right) \right) Ad\left(
\exp \left( \varepsilon _{2}X_{2}\right) \right) \mathbf{X}
\end{equation*}%
becomes%
\begin{equation*}
\mathbf{X}^{\prime }=a_{1}X_{1}+a_{5}X_{5}
\end{equation*}%
for specific values of the parameters $\varepsilon _{2},~\varepsilon _{3}$
and $\varepsilon _{4}$.

Case 2: For $a_{1}a_{5}=0$, there are three subcases, (a) $a_{1}=0$,~$%
a_{5}\neq 0$;~(b) $a_{1}\neq 0$,~$a_{5}=0$ and (c) $a_{1}=a_{5}=0$.

Case 2a: For $a_{1}=0$ and $a_{2}\neq 0$, and following the steps as before
we find the optimal system $X_{5}$ where $\gamma \neq 1$. In the limit $%
\gamma =1$, the generic optimal system is $a_{3}X_{3}+a_{4}X_{4}+X_{5}$.

Case 2b: For $a_{1}=0$ and $a_{1}\neq 0$ the optimal system is derived
\begin{equation*}
\mathbf{X}^{\prime }=Ad\left( \exp \left( \varepsilon _{4}X_{4}\right)
\right) Ad\left( \exp \left( \varepsilon _{3}X_{3}\right) \right) X
\end{equation*}%
which for specific values of $\varepsilon _{3}$ and $\varepsilon _{4}$ is
simplified as%
\begin{equation*}
\mathbf{X}^{\prime }=a_{1}X_{1}+a_{2}X_{2}
\end{equation*}%
Parameter $a_{2}$ is not an invariant hence, it can be zero too. Hence, the
two optimal systems are~$a_{1}X_{1}+a_{2}X_{2}$ and $X_{1}$.

Case 2c: For $a_{1}=a_{5}=0$, we calculate the generic optimal systems~$%
a_{2}X_{2}+a_{3}X_{3}+a_{4}X_{4}$.

Hence, the one-dimensional optimal systems for $\gamma \neq 1$%
\begin{eqnarray*}
&&X_{1}~,~X_{2}~,~X_{5}~,~aX_{1}+X_{2}~, \\
&&~\alpha X_{1}+X_{5}~,~a_{2}X_{2}+a_{3}X_{3}+a_{4}X_{4}~
\end{eqnarray*}%
and~for $\gamma =1$%
\begin{eqnarray*}
&&X_{1}~,~X_{2}~~~,~aX_{1}+X_{2}~,~\alpha X_{1}+X_{5}~, \\
&&~a_{2}X_{2}+a_{3}X_{3}+a_{4}X_{4}~,~a_{3}X_{3}+a_{4}X_{4}+X_{5}
\end{eqnarray*}

There is a difference in the number of one-dimensional optimal systems which
depends on the parameter $\gamma $, that is expected because the structure
of the Lie algebra changes.

We proceed our analysis by applying the Lie symmetries to reduce the system
of PDEs into a system of ODEs and solve the resulting ODEs by applying the
method of Lie symmetries.

\begin{table}[tbp] \centering%
\caption{Commutators of the admitted Lie point symmetries by system
\ref{sw.04}-\ref{sw.05}}%
\begin{tabular}{c|ccccc}
\hline\hline
$\left[ ~,~\right] $ & $X_{1}$ & $X_{2}$ & $X_{3}$ & $X_{4}$ & $X_{5}$ \\
\hline
$X_{1}$ & $0$ & $0$ & $-X_{4}$ & $X_{3}$ & $0$ \\
$X_{2}$ & $0$ & $0$ & $0$ & $0$ & $\left( \gamma +1\right) X_{2}$ \\
$X_{3}$ & $X_{4}$ & $0$ & $0$ & $0$ & $\left( \gamma -1\right) X_{3}$ \\
$X_{4}$ & $-X_{3}$ & $0$ & $0$ & $0$ & $\left( \gamma -1\right) X_{4}$ \\
$X_{5}$ & $0$ & $-\left( \gamma +1\right) X_{2}$ & $-\left( \gamma -1\right)
X_{3}$ & $-\left( \gamma -1\right) X_{4}$ & $0$ \\ \hline\hline
\end{tabular}%
\label{tabl1}%
\end{table}%

\begin{table}[tbp] \centering%
\caption{Adjoint representation for the Lie point symmetries of the system
\ref{sw.04}-\ref{sw.05}}%
\begin{tabular}{c|ccccc}
\hline\hline
$Ad\left( \exp \left( \varepsilon X_{i}\right) \right) X_{j}$ & $X_{1}$ & $%
X_{2}$ & $X_{3}$ & $X_{4}$ & $X_{5}$ \\ \hline
$X_{1}$ & $X_{1}$ & $X_{2}$ & $\cos \left( \varepsilon \right) X_{3}+\sin
\left( \varepsilon \right) X_{4}$ & $\cos \left( \varepsilon \right)
X_{4}-\sin \left( \varepsilon \right) X_{3}$ & $X_{5}$ \\
$X_{2}$ & $X_{1}$ & $X_{2}$ & $X_{3}$ & $X_{4}$ & $X_{5}-\varepsilon \left(
\gamma +1\right) X_{2}$ \\
$X_{3}$ & $X_{1}-\varepsilon X_{4}$ & $X_{2}$ & $X_{3}$ & $X_{4}$ & $%
X_{5}-\varepsilon \left( \gamma -1\right) X_{3}$ \\
$X_{4}$ & $X_{1}+\varepsilon X_{3}$ & $X_{2}$ & $X_{3}$ & $X_{4}$ & $%
X_{5}-\varepsilon \left( \gamma -1\right) X_{4}$ \\
$X_{5}$ & $X_{1}$ & $e^{\varepsilon \left( \gamma +1\right) }X_{2}$ & $%
e^{\varepsilon \left( \gamma -1\right) }X_{3}$ & $e^{\varepsilon \left(
\gamma -1\right) }X_{4}$ & $X_{5}$ \\ \hline\hline
\end{tabular}%
\label{tabl2}%
\end{table}%

\subsection{Static solution}

The application of the symmetry vector $X_{1}$ in (\ref{sw.04})-(\ref{sw.05}%
) provides with the static solution $h=H\left( t_{0},x\right) $ and $%
v=V\left( t_{0},x\right) $. The system of PDEs reduce to one first-order ODE
\begin{equation}
\frac{1}{\gamma }\left( H^{\gamma }\right) _{,x}-V=0,
\end{equation}%
which provides a constraint condition between the velocity $v$ and the
height $h$.

\subsection{Point solution}

The application of the symmetry vector $X_{2}$ provides with the
time-dependent solution in a specific point, i.e. $h=H\left( t,x_{0}\right) $
and $v=V\left( t,x_{0}\right) $. The resulting system provides $H\left(
t,x_{0}\right) =H_{0}$ and the second-order ODE%
\begin{equation}
V_{tt}+V=0.  \label{sw.06}
\end{equation}%
The later equation is nothing else than the oscillator which admits eight
Lie point symmetries and it is maximally symmetric. The Lie symmetries $%
X_{1},X_{3}$ and $X_{4}$ are inherit symmetries, while the rest five Lie
point symmetries are Type II symmetries. The exact solution of equation (\ref%
{sw.06}) is%
\begin{equation}
V\left( t,x_{0}\right) =V_{1}\cos \left( t\right) +V_{2}\sin \left( t\right)
.  \label{sw.07}
\end{equation}

\subsection{Travel-wave solution}

The linear combination of $X_{1}+cX_{2}$ provides travel-wave solutions $%
h=H\left( x-ct\right) ,~v=V\left( x-ct\right) $ \ where parameter $c$
describes the the wave speed. The reduced system is%
\begin{eqnarray}
V_{\xi \xi }-H^{-2}H_{\xi } &=&0,  \label{sw.08} \\
c^{2}V_{\xi \xi }-H^{\gamma -1}H_{\xi }+V &=&0.  \label{sw.09}
\end{eqnarray}%
in which the new independent parameter $\xi $ is defined as $\xi =x-ct$.

From equation (\ref{sw.08}) we derive
\begin{equation}
H^{-1}=H_{0}-V_{\xi },  \label{sw.10}
\end{equation}%
where by substitute in (\ref{sw.09}) it follows
\begin{equation}
\left( c^{2}-\left( H_{0}-V_{\xi }\right) ^{-\gamma -1}\right) V_{\xi \xi
}+V=0.  \label{sw.11}
\end{equation}

The latter equation admits only one Lie point symmetry for $\gamma \succeq 1$%
, the autonomous symmetry $\partial _{\xi }.$ Recall that for $\gamma =-1,$
equation (\ref{sw.11}) becomes a maximally symmetric equation but such value
for parameter $\gamma $ is not physical accepted.

Application of the differential invariants of the autonomous symmetry vector
$\partial _{\xi }$ in (\ref{sw.11}) lead to the$~$nonlinear first-order ODE%
\begin{equation}
\frac{w}{\left( H_{0}-w\right) }\frac{dw}{dz}=z\left( \left( H_{0}-w\right)
^{-\gamma }-c^{2}\left( H_{0}-w\right) \right) ^{-1},  \label{sw.12}
\end{equation}%
with solution%
\begin{equation}
\gamma \left( \gamma -1\right) \left( z^{2}+c^{2}w^{2}+w_{0}\right) +\left(
H_{0}-w\right) ^{-\gamma +1}=0,  \label{sw.13}
\end{equation}%
where the new variables $\left\{ z,w\left( z\right) \right\} ~$are defined
as $z=V\left( \xi \right) $ and $w\left( z\right) =V_{\xi }$.

In the simplest case where the integration constant $H_{0}$ vanishes, and $%
\gamma =1$, the generic solution is given in terms of the Lambert function%
\begin{equation}
\ln \left( w\left( z\right) \right) =-\frac{1}{2}W\left( -c^{2}\exp \left(
z^{2}+2w_{0}\right) \right) +\frac{z^{2}}{2}+w_{0}.  \label{sw.14}
\end{equation}

\begin{figure}[tbp]
\includegraphics[width=0.9\textwidth]{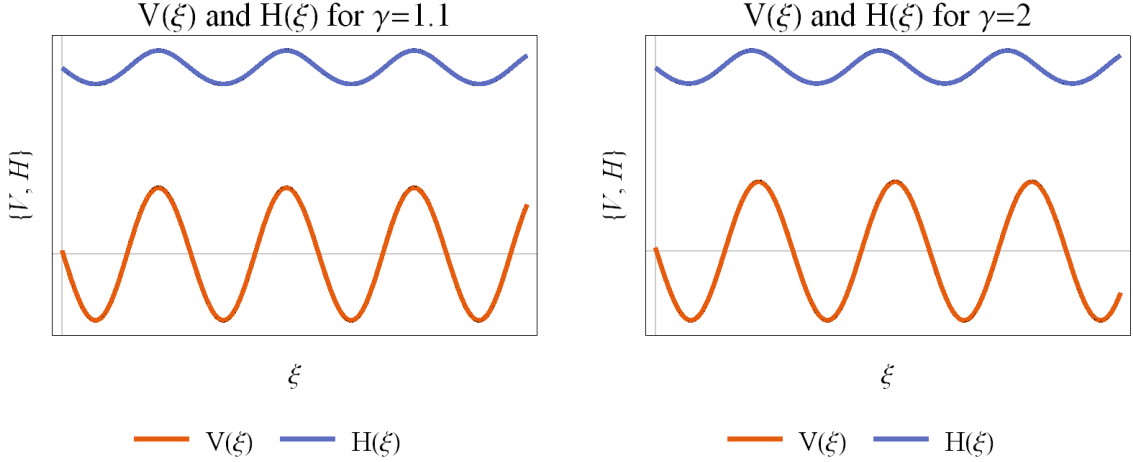}
\caption{Qualitative evolution of the functions $V\left( \protect\xi \right)
$ and $H\left( \protect\xi \right) $ provided by the numerical simulation of
the nonlinear differential equation (\protect\ref{sw.11}). The plots are for
$c=1$ and $H_{0}=2$ and initial conditions $V\left( 0\right) =0.01$,~$V_{%
\protect\xi }\left( 0\right) =-0.2$. Left figure is for $\protect\gamma =1.1$
while right figure is for $\protect\gamma =2$. We observe that $V\left(
\protect\xi \right) $ and $H\left( \protect\xi \right) $ are periodic
functions and have similar behaviour, the different values of $\protect%
\gamma $ changes only the frequency of the oscillations. However, as we
decreased the initial value $V_{\protect\xi }\left( 0\right) $ in values
where $c^{2}-\left( H_{0}-V_{\protect\xi }\right) ^{-\protect\gamma %
-1}\simeq 1$ then the numerical simulation provided singular behaviour for
the $V\left( \protect\xi \right) $ which correspond to a shock. \ }
\label{fig1}
\end{figure}

In Fig. \ref{fig1} we present a numerical simulation of the $H\left( \xi
\right) $ and $V\left( \xi \right) $ as they provided by the differential
equation (\ref{sw.11}). The plots which are presented are for $\gamma =1.1$
and $\gamma =2$. From the figure we observe a travelling-wave solution for $%
V\left( \xi \right) $, and for the variable $H\left( \xi \right) $.

\subsection{Scaling solution}

The Lie invariants of the scaling symmetry vector $X_{5}$ are
\begin{equation}
h=H\left( t\right) x^{\frac{2}{\gamma +1}}~,~v=V\left( t\right) x^{\frac{%
\gamma -1}{\gamma +1}}.  \label{sw.15}
\end{equation}%
Hence, the reduced system consists by two second-order ODEs
\begin{eqnarray}
\frac{\gamma -1}{\gamma +1}V_{t}+H^{-2}H_{t} &=&0,  \label{sw.16} \\
V_{tt}-\frac{2}{\gamma +1}H^{\gamma }+V &=&0.  \label{sw.17}
\end{eqnarray}

From (\ref{sw.16}) and for $\gamma >1~$we find
\begin{equation}
V\left( t\right) =-\frac{\gamma +1}{\gamma -1}H^{-1}+V_{0},  \label{sw.18}
\end{equation}%
and replacing in (\ref{sw.17}) we end up with one second-order ODE with
dependent variable the $H\left( t\right) $, i.e.,
\begin{equation}
Z_{tt}+Z-\frac{\left( \gamma ^{2}-1\right) }{\left( \gamma +1\right) ^{2}}%
V_{0}+2\frac{\left( \gamma -1\right) }{\left( \gamma +1\right) ^{2}}%
Z^{-\gamma }=0  \label{sw.19}
\end{equation}%
where we have replaced $H=Z^{-1}$ in order to simplify the form of the
differential equation.

For arbitrary parameter $\gamma $, equation (\ref{sw.19}) admits only the
autonomous symmetry vector $\partial _{t}$. In the special case where $%
V_{0}=0$ and $\gamma =3$ equation (\ref{sw.19}) is invariant under the $%
sl\left( 3,R\right) $ Lie algebra and reduce to the Ermakov-Pinney equation
\cite{Ermakov,Pinney}. We proceed with the application of the autonomous
vector field.

The Lie invariants of the autonomous symmetry are $z=Z$ and $w=Z_{t}$,
hence, equation (\ref{sw.19}) reduces to the first-order ODE%
\begin{equation}
\frac{d}{dz}\left( \frac{w^{2}}{2}\right) =\frac{\left( \gamma ^{2}-1\right)
}{\left( \gamma +1\right) ^{2}}V_{0}-2\frac{\left( \gamma -1\right) }{\left(
\gamma +1\right) ^{2}}z^{-\gamma }-z.  \label{sw.20}
\end{equation}%
with solution%
\begin{equation}
w\left( z\right) ^{2}=2\frac{\left( \gamma ^{2}-1\right) }{\left( \gamma
+1\right) ^{2}}V_{0}z-\frac{4}{\left( \gamma +1\right) ^{3}}z^{1-\gamma
}-z^{2}.  \label{sw.21}
\end{equation}

For $\gamma =1$, the Lie invariants of the scaling symmetry $X_{5}$ are $%
h=H\left( t\right) x~,~v=V\left( t\right) $, where the reduced system gives,
$H_{t}=0$, i.e. $H=H_{0}$ and $V_{tt}+V=0$.

\subsection{Reduction with the vector fields $X_{3}~\&~X_{4}$}

The existence of the two symmetry vectors $X_{3}$, $X_{4}$ or of the more
general symmetry~$\Gamma =\cos \left( t+t_{0}\right) \partial _{v}$,
indicates that the system of our consideration (\ref{sw.04})-(\ref{sw.05})
is invariant under the point transformation%
\begin{equation}
v\rightarrow v+\cos \left( t+t_{0}\right) .  \label{sw.22}
\end{equation}

Consequently, by taking any linear combination of the other symmetries with
the vector field $\Gamma $ we determine the same reduced equations, where
the invariants have been transformed according to the rule\ (\ref{sw.22}),
except to the case of point reduction, i.e. $X_{2}$ and scaling solution for
$\gamma =1$, which are the two cases we present in the following lines.

\subsubsection{Reduction with $X_{2}+\protect\beta \Gamma $}

By considering the Lie invariants in (\ref{sw.04})-(\ref{sw.05}) of the
symmetry vector $X_{2}+\beta \Gamma $ we find $h=H\left( t\right) $ and $%
v=\beta x\cos \left( t+t_{0}\right) +V\left( t\right) $ where%
\begin{eqnarray}
H^{-2}H_{t}+\beta \sin \left( t+t_{0}\right) &=&0,  \label{sw.23} \\
V_{tt}+V &=&0.  \label{sw.24}
\end{eqnarray}
Therefore, the main difference with the point reduction solution is that the
height $h$ is not a constant anymore, and it is given by%
\begin{equation}
H\left( t\right) =-\left( H_{0}-\beta \cos \left( t+t_{0}\right) \right)
^{-1},  \label{sw.25}
\end{equation}%
while now
\begin{equation}
v=\beta x\cos \left( t+t_{0}\right) +V_{1}\cos \left( t+t_{1}\right)
\label{sw.26}
\end{equation}

It is important here to mention that in order to avoid singular behaviour in
finite time, then the integration constants $H_{0}\neq 0$ while $\beta
<H_{0} $. The latter solution provide a linear spread of the velocities in
the space.

\subsubsection{Reduction with $X_{5}+\protect\beta \Gamma ~$for $\protect%
\gamma =1$}

The case of scaling solutions with $\gamma =1~$when we consider the symmetry
vector $\Gamma $ in the reduction is totally different from the case before.
Indeed, the Lie invariants are
\begin{equation}
h=H\left( t\right) x~,~v=\frac{\beta }{2}\cos \left( t+t_{0}\right) \ln
x+V\left( t\right) ,  \label{sw.27}
\end{equation}%
where $H\left( t\right) $ and $V\left( t\right) $ satisfy the reduced
equations%
\begin{eqnarray}
2H^{-2}H_{t}+\beta \sin \left( t+t_{0}\right) &=&0,  \label{sw.28} \\
V_{tt}+V-H &=&0.  \label{sw.29}
\end{eqnarray}

From (\ref{sw.28}) it follows
\begin{equation}
H\left( t\right) =\frac{2}{H_{0}-\beta \cos \left( t+t_{0}\right) }
\label{sw.30}
\end{equation}%
where by replacing in (\ref{sw.29}) gives a maximally symmetric second-order
ODE with generic solution%
\begin{eqnarray}
V\left( t\right) &=&V_{1}\cos \left( t+t_{0}+t_{1}\right) +\frac{2}{\beta }%
\left( 2\sin \left( t+t_{0}\right) \arctan \left( \frac{\cos \left(
t+t_{0}\right) -1}{\sin \left( t+t_{0}\right) }\right) -\cos \left(
t+t_{0}\right) \ln \left( \beta \cos \left( t+t_{0}\right) -2H_{0}\right)
\right) +  \notag \\
&&-8\left( \beta \sqrt{4H_{0}^{2}-\beta ^{2}}\right) ^{-1}H_{0}\sin \left(
t+t_{0}\right) \arctan \left( \frac{2H_{0}+\beta }{\sqrt{4H_{0}^{2}-\beta
^{2}}}\frac{\cos \left( t+t_{0}\right) -1}{\sin \left( t+t_{0}\right) }%
\right)  \label{sw.31}
\end{eqnarray}%
while the conditions follows $\beta <H_{0}$ in order to avoid singularities
at finite time.

In contrary with the previous reduction where the evolution of the speeds in
the space is linear, in this case, the speed evolve with a logarithmic
expansion which provide an initial singularity at $x=0$.

\subsection{Reduction with the vector fields $\protect\alpha X_{1}+X_{5}$}

For the vector field $\alpha X_{1}+X_{5}$, where $\alpha $ is nonzero
constant the Lie invariants are derived to be
\begin{equation}
\sigma =xe^{-\frac{\gamma +1}{\alpha }t}~,~h=H\left( \sigma \right) e^{\frac{%
2}{\alpha }t}~,~v=V\left( \sigma \right) e^{\frac{\gamma -1}{\alpha }t},
\label{sw.40}
\end{equation}%
from where by replacing in (\ref{sw.04})-(\ref{sw.05}) we find the reduced
system%
\begin{equation}
\sigma H^{2}\left( \gamma ^{2}-1\right) V_{\sigma }-\left( \left( \gamma
-1\right) ^{2}+\alpha ^{2}\right) H^{2}V-\left( \gamma +1\right) ^{2}\sigma
^{2}H_{\sigma }+\alpha ^{2}H^{\gamma +1}H_{\sigma }+2\sigma H=0,
\label{sw.40a}
\end{equation}%
\begin{equation}
\left( \gamma +1\right) \sigma H^{2}V_{\sigma \sigma }-\left( \gamma
+1\right) \sigma H_{\sigma }+2H+2H^{2}V_{\sigma }=0.  \label{sw.40b}
\end{equation}

An exact solution for the later system can be calculated by assuming a
power-law behaviour for the functions $H$ and $V$. Indeed we find the
special solution%
\begin{equation}
V\left( \sigma \right) =V_{0}\sigma ^{\frac{\gamma -1}{\gamma +1}}~,~H\left(
\sigma \right) =H_{0}\sigma ^{\frac{2}{\gamma +1}}  \label{sw.40c}
\end{equation}%
with constraint equation%
\begin{equation}
V_{0}\left( \gamma +1\right) -2H_{0}^{\gamma }=0.  \label{sw.40d}
\end{equation}

In the special case of $\gamma =1$, system (\ref{sw.40a}), (\ref{sw.40b}) is
simplified as follows%
\begin{equation}
\alpha ^{2}H^{2}V-4\sigma H+\left( 4\sigma ^{2}-\alpha ^{2}\right) H_{\sigma
}=0,  \label{sw.41}
\end{equation}%
\begin{equation}
-4\sigma ^{2}V_{\sigma \sigma }-4\sigma V_{\sigma }+\alpha ^{2}\left(
H_{\sigma }-V\right) =0.  \label{sw.42}
\end{equation}%
where now the special solution (\ref{sw.40c}) becomes~$H\left( \sigma
\right) =H_{0}\sigma ,~V\left( \sigma \right) =H_{0}$.

From the system (\ref{sw.41}), (\ref{sw.42}) we can identify the
second-order ODE%
\begin{equation}
Z_{\lambda \lambda }+\frac{\alpha ^{2}\lambda ^{3}+24Z^{\frac{1}{2}}}{%
\lambda \left( \alpha ^{2}\lambda ^{2}-4\right) Z^{\frac{1}{2}}}Z_{\lambda }+%
\frac{2\left( \alpha ^{2}\lambda ^{2}+24Z\right) }{\lambda ^{2}\left( \alpha
^{2}\lambda ^{2}-4\right) }=0.  \label{sw.43}
\end{equation}%
where we have replaced $\lambda =\frac{H\left( \sigma \right) }{\sigma }$
and $Z\left( \lambda \right) =\left( \sigma ^{-1}\left( H\left( \sigma
\right) -\sigma H_{\sigma }\right) \right) ^{\frac{1}{2}}.~$The latter
second-order ODE can be integrated and reduced to the following first-order
ODE
\begin{equation}
\left( \alpha ^{2}\lambda ^{2}-4\right) Z_{\lambda }+\frac{2\alpha \lambda
^{2}\left( 1+\lambda \sqrt{Z}\right) +16Z}{\lambda }+Z_{0}\lambda ^{2}=0.
\label{sw.44}
\end{equation}

\section{Conclusions}

\label{sec4}

In this work we studied a system of nonlinear PDEs which describe rotating
shallow-water with the method of Lie symmetries. More specifically, we
determine the Lie symmetries for the system (\ref{sw.04})-(\ref{sw.05}) and
we found that the system of PDEs is invariant under a five dimensional Lie
algebra. The admitted Lie symmetries form the $\left\{ 2A_{1}\oplus
_{s}2A_{1}\right\} \oplus _{s}A_{1}$ Lie algebra in the
Morozov-Mubarakzyanov Classification Scheme for the parameter $\gamma >1$ or
the $2A_{1}\oplus _{s}3A_{1}$ Lie algebra when $\gamma =1$. That difference
in the admitted Lie algebra between the two cases $\gamma =1$ and $\gamma
\neq 1$ are observable in the application of Lie symmetries and more
specifically in the reduction process.

Indeed for any symmetry vector we considered the application of the
zero-order Lie invariants and we rewrote the system of PDEs into a system
ODEs which we were able to solve them explicitly in all cases by using the
Lie symmetry analysis. Another important feature of the Lie symmetries is
that we can transform solutions into solutions after the application of the
invariant 1PPT.\

From the Lie symmetry vectors $X_{1}-X_{5}$ of the system (\ref{sw.04})-(\ref%
{sw.05}) we determine the generic 1PPT to be%
\begin{eqnarray*}
\bar{t} &=&c_{1}\left( t+\varepsilon \right) \\
\bar{x} &=&c_{2}\left( x+\varepsilon \right) +c_{5}e^{\left( \gamma
+1\right) \varepsilon }x \\
\bar{h} &=&c_{3}h+c_{5}e^{2\varepsilon }h \\
\bar{v} &=&c_{4}\left( v+\varepsilon \cos \left( t+t_{0}\right) \right)
+c_{5}e^{\left( \gamma -1\right) \varepsilon }v
\end{eqnarray*}

It is important to mention that someone could started the present analysis
by studying the original three-dimensional first-order differential
equations (\ref{sw.01})-(\ref{sw.03}). Either in that approach the results
of the analysis will not change, except from that the point transformation
is defined in the space of variables $J=\left\{ t,x,h,v,u\right\} $. In our
analysis we consider to study the point transformations defined in the space
of variable $\,I=\left\{ t,x,h,v\right\} $. Now by extending the symmetry
vectors obtained in the space $I$ in the (partial-) extension space $%
I^{\prime }=$ $\left\{ t,x,h,v,\frac{\partial v}{\partial t}\right\} $, the
symmetry vectors~$X_{1-5}$ become
\begin{eqnarray*}
X_{1} &=&\partial _{t}~+0\partial _{v_{t}},~X_{2}=\partial _{x}+0\partial
_{v_{t}} \\
X_{3} &=&\cos \left( t\right) \partial _{v}-\sin \left( t\right) \partial
_{v_{t}}~,~X_{4}=\sin \left( t\right) \partial _{v}+\cos \left( t\right)
\partial _{v_{t}} \\
X_{5} &=&\left( \gamma +1\right) x\partial _{x}+\left( \gamma -1\right)
v\partial _{v}+2h\partial _{h}+\left( \gamma -1\right) v_{t}\partial _{v_{t}}
\end{eqnarray*}%
where by replacing $u=-v_{t}$. \ are the Lie point symmetries of the
original system (\ref{sw.01})-(\ref{sw.03}).

We conclude that with the application of Lie symmetries we were able to
prove the existence of solutions for the rotating shallow wave system (\ref%
{sw.04})-(\ref{sw.05}). Another important observation is that the reduced
differential equations were reduced into well-known first-order ODEs.
Finally, we proved the existence of travel-wave and scaling solutions.

\bigskip

\textbf{Acknowledgements}

\textit{The author acknowledges the partial financial support of FONDECYT
grant no. 3160121 and thanks the University of Athens for the hospitality
provided.}

\appendix

\section{Lie symmetry analysis in the Euler coordinates}

The dynamical system (\ref{sw.01})-(\ref{sw.03}) in the Euler coordinates is
written as \cite{ref1}
\begin{eqnarray}
h_{t}+\left( hu\right) _{x} &=&0,  \label{sw.32} \\
\left( hu\right) _{t}+\left( hu^{2}+\frac{1}{\gamma }h^{\gamma }\right)
_{x}-hv &=&0,  \label{sw.33} \\
\left( hv\right) _{t}+\left( huv\right) _{x}+hu &=&0.  \label{sw.34}
\end{eqnarray}

The symmetry analysis provide the same results in the Euler coordinates. The
applications of the symmetry condition (\ref{ls.05a}) gives the symmetry
vector fields
\begin{eqnarray*}
Y_{1} &=&\partial _{t}~\ ,~~Y_{2}=\partial _{x}~~,~Y_{3}=\sin \left(
t\right) \left( \partial _{x}-\partial _{v}\right) +\cos \left( t\right)
\partial _{u}~,~ \\
Y_{4} &=&\cos \left( t\right) \left( \partial _{x}-\partial _{v}\right)
-\sin \left( t\right) \partial _{u}~,~Y_{5}=\left( \gamma -1\right) \left(
x\partial _{x}+u\partial _{u}+v\partial _{v}\right) +2h\partial _{h}.
\end{eqnarray*}

\begin{figure}[tbp]
\includegraphics[width=0.9\textwidth]{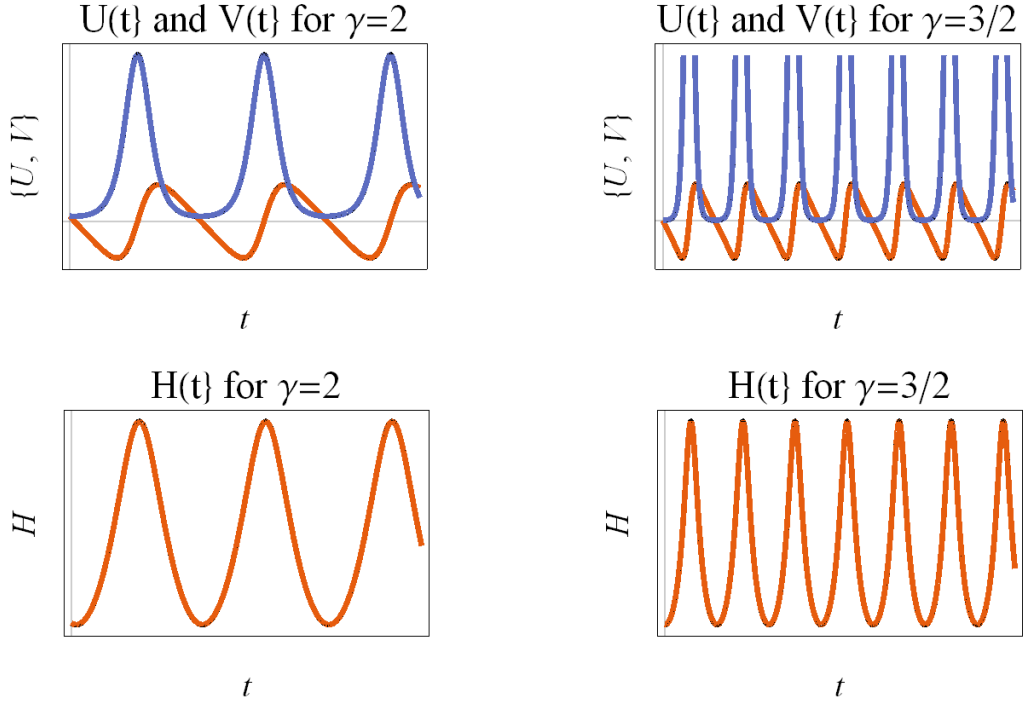}
\caption{Qualitative evolution of the parameters $H\left( t\right) ,~U\left(
t\right) $ and $V\left( t\right) $ given by the system (\protect\ref{sw.36}%
)-(\protect\ref{sw.39}) for $\protect\gamma =2$ and $\protect\gamma =3/2.$
For initial conditions we assumed $H\left( 0\right) =0.02,~U\left( 0\right)
=0.01$ and $V\left( 0\right) =0.01$. The numerical solutions provide
oscillations for the dynamical parameters.}
\label{fig2}
\end{figure}

The latter symmetry vectors form the same Lie algebras with that of system (%
\ref{sw.01})-(\ref{sw.03}). Because they are in a different representation
with the vector fields $X_{1}-X_{5}$, the invariant functions are different.
Vector fields $Y^{1}$ and $Y^{2}$ provide static and stationary solutions
while the linear combination $cY_{1}+Y_{2}~$gives the similarity solution
which describe traveling waves. For $\gamma \neq 1,~$the vector field $Y_{5}$
provide the scaling invariants%
\begin{equation}
h=H\left( t\right) x^{\frac{2}{\gamma -1}}~,~u=U\left( t\right)
x~,~v=V\left( t\right) x
\end{equation}%
where functions $H\left( t\right) ,~U\left( t\right) $ and $V\left( t\right)
$ satisfy the following system of first-order ODEs
\begin{eqnarray}
\left( \gamma -1\right) H_{t}+\left( \gamma +1\right) HU &=&0,  \label{sw.36}
\\
\left( \gamma -1\right) \left( \left( HU\right) _{t}-HV\right) +2\gamma
\left( HU^{2}+H^{\gamma }\right) &=&0,  \label{sw.37} \\
\left( \gamma -1\right) \left( HV\right) _{t}+\left( \gamma -1\right)
UV+2\gamma HUV &=&0.  \label{sw.39}
\end{eqnarray}%
The latter system can be easily integrated, however its general solution it
is not given by a closed-form expression. Numerical simulations of the
latter system are presented in Fig. \ref{fig2} for two values of the
parameter $\gamma .~$More specifically for $\gamma =2$ and $\gamma =\frac{3}{%
2}$. From the figures we observe periodic behaviour for the dynamical
parameters of the dynamical system.

Finally, the symmetry vector $aY_{2}+Y_{3}$ provide the generic solution
\begin{equation}
h\left( t,x\right) =h_{0}\left( a+\sin \left( t\right) \right)
^{-1}~,~u\left( t,x\right) =\frac{\cos \left( t\right) }{\left( a+\sin
\left( t\right) \right) }x+U\left( t\right) ~,~v\left( t,x\right) =-\frac{%
\sin \left( t\right) }{\left( a+\sin \left( t\right) \right) }x+V\left(
t\right) ,
\end{equation}%
where%
\begin{equation}
U\left( t\right) =\frac{U_{1}a\cos \left( t\right) +U_{2}\left( a\sin \left(
t\right) +1\right) }{a\left( a+\sin \left( t\right) \right) }~,~V\left(
t\right) =\frac{U_{2}\cos \left( t\right) -U_{1}\sin \left( t\right) }{%
\left( a+\sin \left( t\right) \right) }.
\end{equation}


\begin{thebibliography}{99}
\bibitem{Ovsi} L. V. Ovsiannikov, Group analysis of differential equations,
Academic Press, New York, (1982)

\bibitem{harrison} B.K. Harrison, Sigma 1, 001 (2005)

\bibitem{refA1} V.A. Baikov, A.V. Gladkov and R.J. Wiltshire, J. Phys. A:
Math. Gen. 31, 7483 (1998)

\bibitem{ser1} T.G. Mkhize, K.\ Govinder, S. Moyo and S.V. Meleshko, Appl.
Math. Comput. 301, 25 (2017)

\bibitem{ser2} A. Paliathanasis and P.G.L. Leach, Int. J. Geom. Meth. Mod.
Phys. 13, 1630009 (2016)

\bibitem{nuc1} M.C. Nucci and P.G.L. Leach, J. Math. Anal. Appl. 406, 219
(2013)

\bibitem{nuc2} M.C. Nucci, J. Nonl. Math. Phys. 20, 451 (2013)

\bibitem{noeth} E. Noether, Nachr. d. K\"{o}nig. Gesellsch. d. Wiss. zu G%
\"{o}ttingen, Math-phys. Klasse, 235, (1918) (translated in English by M.A.
Tavel [physics/0503066])

\bibitem{nibr} N.H. Ibragimov, J. Math. Anal. Appl. 333, 311 (2007)

\bibitem{sarlet} W.\ Sarlet and F. Cantrijin, SIAM Review, 23, 467 (1981)

\bibitem{olver} P.J. Olver, Applications of Lie Groups to Differential
Equations, Springer-Verlag, New York, (1993)

\bibitem{kumei} G.W. Bluman and S. Kumei, Symmetries and Differential
Equations, Springer-Verlag, New York, (1989)

\bibitem{ibra} N.H. Ibragimov, CRC Handbook of Lie Group Analysis of
Differential Equations, Volume I: Symmetries, Exact Solutions, and
Conservation Laws, CRS Press LLC, Florida (2000)

\bibitem{serb} S.V. Meleshko, Methods for Constructing Exact Solutions of
Partial Differential Equations, Springer Science,\ New York (2005)

\bibitem{app1} S.V. Meleshko and V.P. Shapeev, J. Nonl. Math. Phys. 18, 195
(2011)

\bibitem{app2} G.M. Webb and G.P. Zank, J. Math. Phys. A: Math. Theor. 40,
545 (2007)

\bibitem{app3} M.C. Nucci and G. Sanchini, Symmetry 7, 1613 (2015)

\bibitem{app4} A. Paliathanasis, K. Krishnakumar, K.M. Tamizhmani and P.G.L.
Leach, Mathematics 4, 28 (2016)

\bibitem{ap11} X. Xin, Appl. Math. Lett. 55, 63 (2016)

\bibitem{ap12} X. Xin, Acta Phys. Sin. 65, 240202 (2016)

\bibitem{app5} N. Kallinikos and E.\ Meletlidou, J. Phys. A: Math. Theor.
46, 305202 (2013)

\bibitem{app6} S. Jamal and A. Paliathanasis, J. Geom. Phys. 117, 50 (2017)

\bibitem{app7} G.M. Webb, J. Phys A: Math. Gen. 23, 3885 (1990)

\bibitem{app8} P.G.L. Leach, J. Math. Anal. Appl. 348, 487 (2008)

\bibitem{app9} M. Tsamparlis and A. Paliathanasis, J. Phys. A: Math. Theor.
44, 175202 (2011)

\bibitem{app10} M. Tsamparlis and A. Paliathanasis, Symmetry (MDPI) 10, 233
(2018)

\bibitem{ap13} X. Xin, Commun. Theor. Phys. 66, 479 (2016)

\bibitem{ap14} X. Xin, H. Liu, L. Zhang and Z. Wang, Appl. Math. Lett. 88,
132 (2019)

\bibitem{ref1} B. Cheng, P. Qu and C. Xe, SIAM J. Math. Anal. 50, 2486 (2018)

\bibitem{ref2} B. Galperin, H. Nakano, H.-P. Huang, and S. Sukoriansky,
Geoph. Res. Let. 31, L13303 (2004)

\bibitem{ref3} V. Zeitlin, S. B. Medvedev, and R. Plougonven, J. Fluid.
Mech. 481, 269 (2003)

\bibitem{ref4} D.A. Randall, Monthly Weather Review 122, 1371 (1994)

\bibitem{swr02} M. Senthilvelan and M. Lakshmanan, Int. J. Nonl. Mech. 31,
339 (1996)

\bibitem{swr03} S. Szatmari and A. Bihlo, Comm. Nonl. Sci. Num. Sim. 19, 530
(2014)

\bibitem{swr04} A.A. Chesnokov, J. Appl. Mech. Techn. Phys. 49, 737 (2008)

\bibitem{swr05} J.-G. Liu, Z.-F. Zeng, Y. He and G.-P. Ai, Int. J. Nonl.
Sci. Num. Sim. 16, 114 (2013)

\bibitem{swr06} A.A. Chesnokov, Eur. J. Appl. Math. 20, 461 (2009)

\bibitem{swr07} M. Pandey, Int. J. Nonl. Sci. Num. Sim. 16, 93 (2015)

\bibitem{swr08} X. Xin, L. Zhang, Y. Xia and H. Liu, Appl. Math. Lett. 94,
112 (2019)

\bibitem{Morozov58a} Morozov VV (1958), \textit{Izvestia Vysshikh Uchebn
Zavendeni\u{\i} Matematika,} \textbf{5} 161-171

\bibitem{Mubarakzyanov63a} Mubarakzyanov GM (1963), \textit{Izvestia
Vysshikh Uchebn Zavendeni\u{\i} Matematika,} \textbf{32} 114-123

\bibitem{Mubarakzyanov63b} Mubarakzyanov GM (1963), \textit{Izvestia
Vysshikh Uchebn Zavendeni\u{\i} Matematika,} \textbf{34} 99-106

\bibitem{Mubarakzyanov63c} Mubarakzyanov GM (1963), \textit{Izvestia
Vysshikh Uchebn Zavendeni\u{\i} Matematika,} \textbf{35} 104-116

\bibitem{olverb} P.J. Olver, Applications of Lie Groups to Differential
Equations, second edition, Springer-Verlag, New York (1993)

\bibitem{opt1} X. Hu, Y. Li and Y.\ Chen, J. Math. Phys. 56, 053504 (2015)

\bibitem{Ermakov} V. Ermakov, Universita Izvestia Kiev Series III 9 1-25
(1880) (The English version, translated by AO Harin, can be found in
Applicable analysis and Discrete Mathematics)

\bibitem{Pinney} E. Pinney, Proceedings of the American Mathematical Society
1, 681 (1950)
\end{thebibliography}
\end{document}